%% file: paper.tex
\documentclass[11pt,letterpaper]{article}

\input{mesdef.tex}

\usepackage{amssymb}
\usepackage{amsmath}
\usepackage{epsfig}
\usepackage{latexsym}
\usepackage{color}
\usepackage{shadow}
\usepackage{anysize}

\marginsize{2cm}{2cm}{2cm}{2cm}

\newcommand{\mC}{\mathcal{C}}
\newcommand{\+}{\!+\!}
\newcommand{\wtau}{\widetilde{\tau}}
\newcommand{\wDe}{\widetilde{\Delta}}
\newcommand{\wDelta}{\wDe}
\newcommand{\wxi}{\widetilde{\xi}}

\begin{document}
\title{Energy Profile Fluctuations  in Dissipative Nonequilibrium
  Stationary  States}
\author{Jean Farago\footnote{LCP-UMR 8000, Bat. 349, Universit\'e
  Paris-Sud, 91405 Orsay cedex, France. email: farago@lcp.u-psud.fr
  phone (33)-1-69-15-41-95}}
\date{}
\maketitle
\abstract{The exact large deviation function (ldf) for the fluctuations of
  the 
  energy density field is computed for a chain of Ising (or more
  generally Potts) spins
  driven by a zero-temperature (dissipative) Glauber dynamics and
  sustained in a non trivial stationary regime by an arbitrary energy
  injection mechanism at the boundary of the system. It is found that this ldf
  is independent of the dynamical details of the energy injection,
  and that the energy fluctuations, unlike conservative systems in a
  nonequilibrium state,  are not spatially
  correlated in the stationary regime.}

\bigskip

{\bf Keywords:} Large deviations; nonequilibrium stationary states;
  dissipative systems; 
  Glauber dynamics; Potts models.

\newpage

\section{Introduction}

In a series of recent papers, Derrida et al. \cite{seriesofrecentpapers} studied
the nonequilibrium stationary states of exactly solvable models
characterized by \textit{conservative} inner dynamics: particles
diffuse in a one dimensional chain of sites without annihilation, and
are described at the hydrodynamical length- and time-scales by a
Fick's (diffusion) law (with or without a systematic
drift force). The authors succeeded in exactly computing  the
nonequilibrium free energy (associated with a given density profile) which displayed very interesting features:
for instance, the nonequilibrium situation
induces long range correlations which make the free energy non additive. The
authors discovered however an elegant and quite unusual sub-additivity principle whose
generality for other models is still an open question. Their results
also showed that although the second moment of density fluctuations is the
same as previously found with fluctuating hydrodynamics arguments
\cite{spohn}, a quadratic (gaussian) approximation is not sufficient
for higher cumulants.

\bigskip

In this paper, we address the same question of characterizing of
nonequilibrium stationary states (NESS), but for dissipative inner dynamics. Such situations are widely encountered: for instance, the
granular matter  continuously shaked by a piston can be considered for
long time scales as being in a nonequilibrium stationary state; the
velocity field of a turbulent flow is also subjected to a dissipative
(Navier-Stokes) equation and can be sustained in a stationary state as
well. These dissipative systems are altogether different from
conservative ones, since no concept of thermodynamical equilibrium can
be applied to them: in absence of energy input, the full rest is
the ultimate state of the system; as a result, no pertubative strategy
can be deployed to describe their physics, and in fact the
non equilibrium stationary states of dissipative systems are not well
known (see for instance \cite{aumaitrefauve,moon} and references therein). 

In particular, at the light of the recent works aforementioned,
natural questions arise for dissipative NESS: are they similarly
characterized by a high level of correlation ? are the fluctuations
to some extent insensitive to the details of the inner dynamics
and/or the injection mechanism ? is the notion of nonequilibrium
``thermodynamical potential'' even relevant ?

To answer these questions, and to develop further the ideas presented in
\cite{jeanlangevin,jeanlangevin2}, we consider here a one-dimensional
dissipative model for which some exact results on the energy
structuration can be extracted: our model is a semi-infinite 1D chain of Ising
spins (easily generalized to a Potts model)  subjected to
zero-temperature Glauber dynamics. These intrinsically dissipative
dynamics are supplemented with an arbitrary (Poissonian to be simple)
flipping process of the first spin as a way to inject energy into the system.

Our results can be summarized as follows: i) the concept of
nonequilibrium potential can be extended via a non homogeneous
space coarsening of this dissipative system; ii) this potential
is \textit{independent} of the nature of the injection mechanism, and
reflects mainly the dynamical inner self organization of the system; iii) this potential
\textit{does not} display correlations, i.e. it is ``additive'', as
soon as a stretching of the space is taken into account; iv) such a
potential cannot be defined for any dissipative system, and the
conditions that system dynamics must fulfill to have a global observable
obeying the large deviation theorem are not clear.

The paper is organized as follows. In a first part we define the model
and its dynamics; then we give some preliminary results concerning the
stationary state and introduce the cumulative energy $E$. Afterwards
we compute the large deviation function of $E$ and the ldf associated
with an energy profile. We end
up with a discussion of the results.

\section{Definition of the model}
The model we consider is a 1-dimensional semi-infinite chain of Ising spins
$\sig_0,\sig_1,\ldots,\sig_{j},\ldots$ (a finite chain could be
considered as well). The spins follow
zero-temperature Glauber dynamics, except for the spin $\sig_0$, which
flips according to a Poisson process of parameter $\la$ : the
probability associated with a flip of spin $j$ from a configuration
$\mC$ between $t$ and $t+dt$
is $w(\mC\rightarrow\mC_j)dt=[1-\sig_j(\sig_{j+1}+\sig_{j-1})/2]\times
dt$ if $j\neq 0$ (the starting configuration is termed $\mC$ and the
$j$-flipped configuration $\mC_j$)  and
$\la dt$ for $\sig_0$. As a result, the master equation describing dynamics in the model is 
\begin{align}\label{masterequation}
  \pa_t
  P(\mC)&=-\sum_{j\geq 0}w(\mC\rightarrow\mC_j)P(\mC)+\sum_{j\geq 0}w(\mC_j\rightarrow\mC)P(\mC_j)\\
&= -\la[P(\mC)-P(\mC_0)]-\sum_{j\geq
  1}[1-\sig_j(\sig_{j+1}+\sig_{j-1})/2]P(\mC)+\sum_{j\geq 1}[1+\sig_j(\sig_{j+1}+\sig_{j-1})/2]P(\mC_j)
\end{align}
In the last equality, the spin values refer obviously to configuration
$\mC$.

\medskip

Inner Glauber dynamics are essentially dissipative: the domain
walls (which are elementary energy excitations) move randomly and
annihilate by pairs when colliding. Thus, in absence of energy
input, any initial condition would eventually relax to a state
characterized by the same value of all spins; in the model considered
here, such an external input is provided by the Poissonian motion of
the spin labeled $0$,  which gives energy to the system when $\sig_0$
flips from $\sig_1$ to $-\sig_1$. As a result,  after a transient, a stationary state
takes place which is described in the
following\footnote{we are aware that the stationary state of the
  infinite system needs in general a diverging time to be
  established, but  this problem can be circumvented by a clever choice
  of the initial conditions}.

\section{The stationary state}

\subsection{Mean injected power}

If the summation $\sum_{\mC}\sig_0\sig_i(\ldots)$ is made on the
dynamical equation \myref{masterequation}, one gets
\begin{align}\label{eqtr1}
  \pa_t\lan\sig_0\sig_i\ran=-2(\la+1)\lan\sig_0\sig_i\ran+\lan\sig_0\sig_{i+1}\ran+\lan\sig_0\sig_{i-1}\ran
\end{align}
When the stationary state is assumed, and according to $\sig_0^2=1$, \myref{eqtr1}
leads to
$\lan\sig_0\sig_i\ran=[1+\la-\sqrt{\la^2+2\la}]^i$ (this correlation
vanishes when $i\rightarrow\infty$).

From this calculation an interesting physical result can be deduced:
the (mean) power $P_\text{inj}$ injected by the spin $0$ inside the system is 
related to $\lan\sig_0\sig_1\ran$, since between $t$ and $t+dt$, the
energy ceded is in average $2\la
dt\times[\text{Prob}(\sig_0=\sig_1)-\text{Prob}(\sig_0=-\sig_1)]$,
whence \cite{werner}
\begin{align}
   P_\text{inj}&=2\la\lan\sig_0\sig_1\ran\\
&=2\la(1+\la-\sqrt{\la^2+2\la})
\end{align}
The injected power is an increasing function of $\la$ which saturates
to $2$ for large $\la$; this value is physically dictated by
internal dynamics, i.e. the ability of the system to diffuse into the
bulk the energy excitations created at the boundary. It is to note
that there is thus no notion of ``optimal'' time scale concerning the
energy injection, which is not evident a priori: one could imagine that
an optimal waiting time could leave domain walls move away from the
boundary.

\subsection{Two-points correlations and energy density profile}

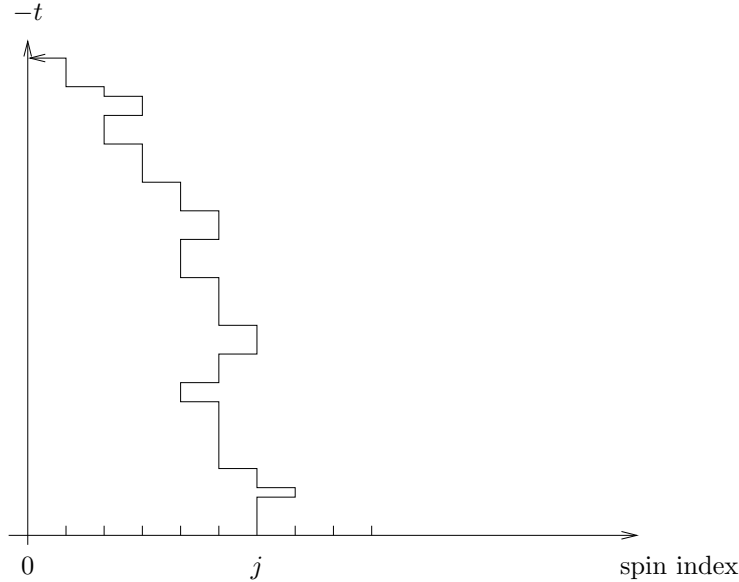
\begin{figure}[h]
  \begin{center}
    \input{tracedback.pstex_t}
\caption{The value of $\sig_j(t)$ can be traced back in time from
  neighbouring spin to neighbouring spin until it reaches $\sig_0$
  whose statistical properties are known.}\label{tracedback}
  \end{center}
\end{figure}

Another interesting quantity describing the stationary state is the
average energy density profile 
\begin{align}
  \lan e_n\ran=1-\lan\sig_n\sig_{n+1}\ran
\end{align}
This quantity is more complicated to obtain. To this end, it is useful
to interpret  Glauber dynamics as coalescing paths dynamics \cite{derridahakimpasquier}: the
update of any spin is equivalent to a  random choice (Poissonian with
parameter 1) of the spin value
among one of its two neighbours, such that the value $\sig_n(t)$ can be
traced back in time from spin to spin until either the time origin or
the boundary $\sig_0$ is reached at a certain spin index or time
(respectively). If the stationary state is studied, the notion of time
origin is irrelevant, and any ``path of constant spin value''
eventually reaches the zeroth spin (see fig. \ref{tracedback}). 
If two spins $\sig_n$ and $\sig_p$ are now monitored, two paths emerge
from $n$ and $p$, are traced back in time, and possibly coalesce if
they meet each other before reaching the boundary. They are statistically
independent, since the flipping processes in two distinct sites are
not correlated. As a result, the average $\lan\sig_n\sig_p\ran$ can be
calculated as follows: consider two random walkers starting at sites $n$ and
$p$. If they meet before touching the site 0, this occurence of
the paths gives a factor 1 in the computation of
$\lan\sig_n\sig_p\ran$. Otherwise, the $n$-path (resp. $p$) arrives at site zero
at time $t_n$ (resp. $t_p$), and this occurence gives a factor
$\lan\sig_0(t_n)\sig_0(t_p)\ran=\exp(-2\la|t_n-t_p|)$, where the average is on the
Poissonian process $\sig_0$. As a result original dynamics are
mapped on dynamics of two random walkers plus dynamics of
$\sig_0$. These considerations can be summarized as
\begin{align}
  \lan\sig_n\sig_p\ran=c_{np}+\int_0^\infty dt_n\int_0^\infty
  dt_p \text{Prob}[t_n,t_p \text{ and } (n,p)] \exp(-2\la|t_n-t_p|)
\end{align}
In this expression, $c_{np}$ is the probability that the walkers
starting at $n$ and $p$ meet each other before reaching the zeroth
site (we will also use $c_{n,p}=1-c_{np}$; generally, a comma is put
between two indices when the associated walkers are supposed to avoid
each other; conversely the absence of comma holds for coalescing walkers); besides $\text{Prob}[t_n,t_p \text{ and }  (n,p)]$
is the probability that the $n$-walker and the  $p$-walker reach site
0 at times $t_n$ and  $t_p$ respectively without having met each other
 beforehand.

\begin{figure}[h]
  \begin{center}
    \includegraphics{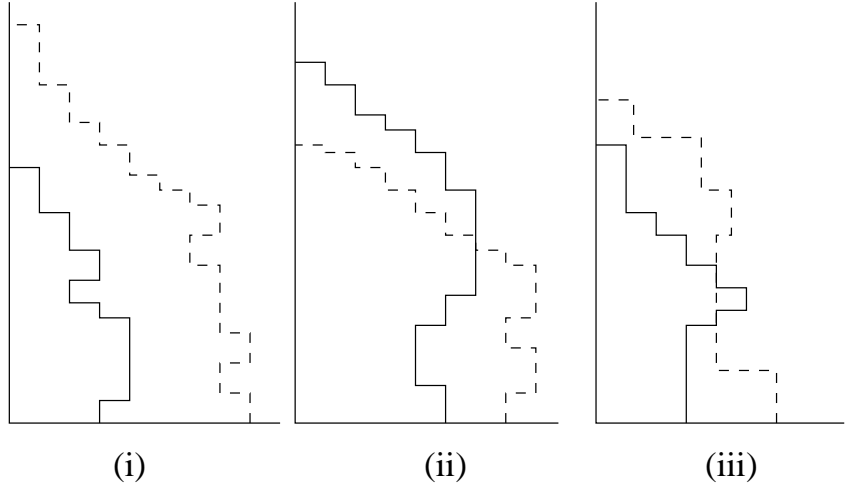}
\caption{Three possible generic situations for two independent walkers}\label{three}
  \end{center}
\end{figure}

\begin{figure}[h]
  \begin{center}
    \input{exchange.pstex_t}
\caption{mapping of type (ii) to type (iii) paths}\label{exchange}
  \end{center}
\end{figure}
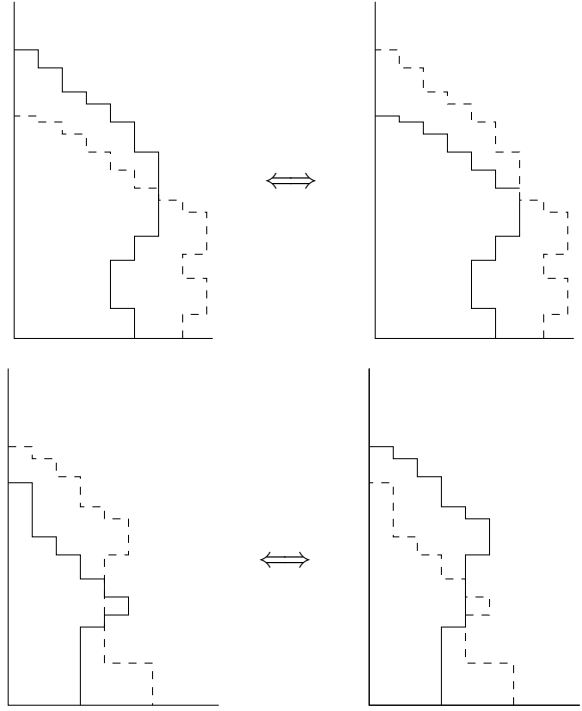

These expressions can be exactly computed. Indeed, consider two
independent random
walkers starting at $n$ and $p$. There are a priori three possible
situations (fig.  \ref{three}): either (i) they do not cross at all, or
(ii) they cross and ``exchange'' their arrival time, or (iii) they
cross without exchanging their arrival time. 
It is easy to
see that each situation of type (ii) can be mapped to a
situation of the type (iii), and vice-versa (fig. \ref{exchange}).
Moreover, the probability is of course conserved by the mapping, since
the two paths are of the same statistical nature. As a result, if $n<p$:
\begin{align}
  c_{n,p}&=\sum_\text{type (i) and (iii) paths}
  \text{Prob(path)}-\sum_\text{type (ii) paths} \text{Prob(path)}\\
&=\int_0^\infty dt_n\int_{t_n}^\infty dt_p\
  [P_n(t_n)P_p(t_p)-P_n(t_p)P_p(t_n)]\\
c_{np}&=2\int_{t_n<t_p}P_n(t_p)P_p(t_n)
\end{align}
where $P_n(\tau)$ is the probability (density) for a walker initially
($t=0$) at $n$  to reach the site 0 \textit{for the first time} at
$t=\tau$. We get, 
for $n\leq p$:
\begin{align}\label{sigisigj}
  \lan\sig_n\sig_p\ran=2\int_0^\infty dt\int_{t}^\infty
  dt'  \ P_n(t')P_p(t)+\int_0^\infty dt\int_{t}^\infty
  dt'  \ [P_n(t)P_p(t')-P_n(t')P_p(t)]\exp(-2\la|t-t'|)
\end{align}
We are going to see that the dominant contribution to
$\lan\sig_n\sig_p\ran$  in the limit $n\rightarrow\infty$ is given by
the first term only. This fact has important consequences on the
structure of the stationary regime.

\bigskip
 
The probability $P_n(t)$ is obtained as a solution of a Brownian
motion with an absorbing boundary at site $0$: the probability
$p(m,t|n)$ that a Brownian walker
starting at site $n$ at $t=0$ be at site $m$ at time $t$ without
having  touched the zeroth site obeys the dynamical equation
$\pa_tp(m|n)=p(m+1|n)+p(m-1|n)-2p(m|n)$ if $m>1$ and $\pa_tp(1|n)=
p(2|n)-2p(1|n)$. The probability density $P_n(t)$ is simply given by
$P_n(t)=p(1,t|n)$ (actually $P_n(t)$ is $p(1,t|n)$ times the transition
rate associated to the jump $1\rightarrow 0$, but this rate were chosen equal
to one).  That $p(m,t|n)=p_f(m-n,t)-p_f(m+n,t)$ (with $p_f$ is the
free-boundary Brownian motion starting at site $0$) is the solution
is directly verified, whence one gets
\begin{align}
P_n(t)&=2\int_0^1 dq\ e^{-4t\sin^2\pi q}\sin(2\pi q)\sin(2\pi qn)
\end{align}
With this expression, one can recast $c_{np}$ in
\begin{align}
  c_{np}=4\int_0^{1/2}dq\ \cot(\pi q)\sin(2\pi q n)\left(2\sin^2\pi
  q+1-2\sqrt{\sin^4\pi q+\sin^2\pi q}\right)^p
\end{align}
An  asymptotic expansion leads to the following result, valid
for $n\rightarrow\infty, p>n$:
\begin{align}\label{delicate}
  c_{np}\simeq\frac{4}{\pi}\text{Atan}\left(\frac{n}{p}\right)
\end{align}

The second term of \myref{sigisigj} can be estimated using the
asymptotic behaviour of $P_n$:
\begin{align}
  P_n(n\tau)&\sur{\simeq}{n\rightarrow\infty}\frac{1}{\sqrt{2\pi
  n}}\frac{1}{\tau(4\tau^2+1)^{1/4}}\exp
  n\left(\sqrt{4\tau^2+1}-2\tau-\log\left[\frac{1}{2\tau}+\sqrt{\frac{1}{4\tau^2}+1}\right]\right)\\
&\sur{\simeq}{\tau\gg 1}\frac{1}{\sqrt{4\pi n\tau^3}}\exp\left(-\frac{n}{4\tau}\right)
\end{align}
A cumbersome computation yields for $n\rightarrow\infty$ and $p>n$:
\begin{align}\label{cumbersome}
  \int_0^\infty dt\int_{t}^\infty
  dt'  \ [P_n(t)P_p(t')-P_n(t')P_p(t)]\exp(-2\la|t-t'|)\simeq \frac{24}{\pi\la^2}\frac{np(p^2-n^2)}{(n^2+p^2)^4}
\end{align}
It is important to remark that in the asymptotic limit
$n\rightarrow\infty$, the dominant contribution to
$\lan\sig_n\sig_p\ran$ is given by the ``coalescing term'' $c_{np}$,
which is \textit{independent} of the value of $\la$ ; moreover the dominant
contribution is (besides the constant 1) \textit{non summable} and
leads to a
logarithmic divergence, whereas the $\la$-dependent corrections are of
order $1/n^5$ (for $n-p=O(1)$) and therefore obviously summable.

\bigskip

These results can be applied to the energy density profile:
\begin{align}\label{moy_en}
  \lan e_n\ran\simeq\frac{2}{\pi n}+O\left(\frac{1}{n^5}\right)
\end{align}
Once again the leading term of the energy density profile is
independent of $\la$ and non summable.

\subsection{Energy density correlations}

Once we have the averaged profile $\lan e_n\ran$, it is natural to
subsequently ask for the correlation function of $e_n$:
\begin{align}\label{correlnrj}
  \lan e_ne_p\ran-\lan e_n\ran\lan e_p\ran=\lan\sig_n\sig_{n+1}\sig_p\sig_{p+1}\ran-\lan \sig_n\sig_{n+1}\ran\lan \sig_p\sig_{p+1}\ran
\end{align}
The 4-point correlator $\lan\sig_n\sig_{n+1}\sig_p\sig_{p+1}\ran$ can
be written
$\lan\sig_n\sig_{n+1}\sig_p\sig_{p+1}\ran=c_{nn+1pp+1}+c_{nn+1,pp+1}+\chi(n,p)$
where $c_{nn+1pp+1}$ is the probability that four walkers starting
at $n$, $n+1$, $p$ and $p+1$ coalesce before reaching the boundary;
$c_{nn+1,pp+1}$ (note the comma) the probability that the walkers
``$n$'' and ``$n+1$'' on one side, walkers ``$p$'' and ``$p+1$''
on the other side coalesce and that the two groups reach the boundary separatly; and $\chi(n,p)$ is the term arising from
situations where  one or several walkers
reach the boundary without coalescence. This $\chi(n,p)$ is obviously $\la$-dependent and is composed of several terms, each corresponding to a
particular scenario for the walkers: with natural notations, $\{n,n\+ 1,p,p\+1\}$,
$\{(n\ n\+1\ p),p\+1\}$, $\{n,(n\+1\ p\ p\+1)\}$, $\{(n\
n\+1),p,p\+1\}$, $\{n,(n\+1\ p),p\+1\}$, $\{n,n\+1,(p\ p\+1)\}$ are the different possible
situations where $\la$ plays a role. Let us consider the first term:
\begin{align}
 \{n,n+1,p,p+1\}=&\int_{t_1<t_2<t_3<t_4} \text{Prob}[t_1,t_2,t_3,t_4 \text{ and
  }(n,n+1,p,p+1)]\lan\sig_0(t_1)\sig_0(t_2)\sig_0(t_3)\sig_0(t_4)\ran
\end{align}
(the probability term refers to four walkers
starting  from the four sites considered and reaching without
coalescence the site $0$ at times $t_1$\ldots$t_4$ respectively).
This  integral must be approximately evaluated; its dominant term
 comes from regions in the $(t_1,t_2,t_3,t_4)$ plane
where $t_1\simeq t_2$ and $t_3\simeq t_4$: in these regions only is
the 4-point correlator in $\sig_0$ significantly different from
zero. As a result,
\begin{align}
  \{n,n+1,p,p+1\}&\sim\la^{-2}\int_{t_1<t_2} \text{Prob}[t_1,t_1+\la^{-1},t_2,t_2+\la^{-1} \text{ and
  }(n,n+1,p,p+1)]\\
&\lesssim \la^{-1}\int_{t_1}\text{Prob}[t_1,t_1+\la^{-1}\text{ and
  }(n,n+1)]\times\la^{-1}\int_{t_2}\text{Prob}[t_2,t_2+\la^{-1}\text{ and
  }(p,p+1)]
\end{align}
But as
\begin{align}
\la^{-1}  \int_{0}^\infty dt\ \text{Prob}[t,t+\la^{-1} \ \text{and}\
    (n,n+1)]&\sim \la^{-1}\int dt P_n(t)P_{n+1}(t+\la^{-1})-P_n(t+\la^{-1})P_{n+1}(t)\\
&\sim \la^{-2}\int_0^\infty dt
    P_n(t)^2\left(\frac{P_{n+1}(t)}{P_n(t)}\right)'\sim \frac{\la^{-2}}{n^5}
\end{align}
we conclude that $\{n,n\+1,p,p\+1\}$ gives a very rapidly decreasing
contribution, and we can notice besides that $\sum_{n<p} \{n,n\+1,p,p\+1\}<\infty$.
The situation is similar if other $\la$-dependent terms are considered. For instance, 
\begin{align}
  \{(n\ n\+1\ p),p\+1\}&\sim \la^{-1} \int_0^\infty dt \text{Prob}[t,t+\la^{-1} \text{ and
  }((n\ n\+1\ p),p\+1)]\\
&\lesssim \la^{-1} \int_0^\infty dt \text{Prob}[t,t+\la^{-1} \text{ and
  }(n,p\+1)]\sim \la^{-2}pn\frac{p^2-n^2}{(p^2+n^2)^4}
\end{align}
(note the similarity between this approximation and the exact limit of
\myref{cumbersome}). Once again this term is very small and verifies $\sum_{n<p} \{(n\
n\+1\ p),p\+1\}<\infty$.

\medskip

A special attention must be paid to the term  $\{n,n\+1,(p\
p\+1)\}$. Actually it is not summable:
$\sum_{n<p<N}\{n,n\+1,(p\
p\+1)\} $ diverges with a contribution $O(N)$ and another $O(\log N)$. But
this term
 is partially ``disconnected'' and these divergences are exactly balanced in
\myref{correlnrj} by a term coming from $\lan \sig_n\sig_{n+1}\ran\lan
\sig_p\sig_{p+1}\ran=(c_{n\mspace{1mu}n\+1}+\{n,n\+1\})(c_{p\mspace{1mu}p\+1}+\{p,p\+1\})$:
actually $\{n,n\+1\}c_{p\mspace{1mu}p\+1}$ is the counterterm which
kills these ``hybrid'' divergences.

\bigskip

So, the correlation function of the energy can be recast as
\begin{align}
   \lan e_ne_p\ran-\lan e_n\ran\lan e_p\ran&=c_{n\mspace{1mu}n\+1\mspace{1mu}p\mspace{1mu}p\+1}+c_{n\mspace{1mu}n\+1,p\mspace{1mu}p\+1}-1+c_{n,n+1}+c_{p,p+1}-c_{n,n+1}c_{p,p+1}+\widetilde{\chi}(n,p)
\end{align}
where $\widetilde{\chi}(n,p)$ depends on $\la$ and verifies
$\sum_{n<p}\widetilde{\chi}(n,p)<\infty$. We will see in the following that
\begin{align}\begin{split}
  c_{n\mspace{1mu}n\+1\mspace{1mu}p\mspace{1mu}p\+1}+c_{n\mspace{1mu}n\+1,p\mspace{1mu}p\+1}=1-&c_{n,p\+1}+c_{n,p}+c_{n\+1,p\+1}-c_{n,n\+1}-c_{n\+1,p}-c_{p,p\+1}\\&+c_{n,n\+1}c_{p,p\+1}+c_{n,p\+1}c_{n\+1,p}-c_{n,p}c_{n\+1,p\+1}
	     \end{split}
\end{align}
whence finally
\begin{align}
  \lan e_ne_p\ran-\lan e_n\ran\lan e_p\ran&=-c_{n,p\+1}+c_{n,p}+c_{n\+1,p\+1}-c_{n\+1,p}+c_{n,p\+1}c_{n\+1,p}-c_{n,p}c_{n\+1,p\+1}+\widetilde{\chi}(n,p)
\end{align}
Local correlations are then easily deduced from \myref{delicate}:
\begin{align}
  \lan e_ne_{n+k}\ran-\lan e_n\ran\lan
  e_{n+k}\ran&\sur{\simeq}{n\text{ large },
  k\rightarrow\infty}-\frac{32n(n+1)(2n+1)}{3\pi^2}\frac{1}{k^5}\\
\lan e_ne_{nk}\ran-\lan e_n\ran\lan
  e_{nk}\ran&\sur{\simeq}{n\text{ large }, k\rightarrow\infty}-\frac{32(n+1)(2n+1)}{3\pi^2}\frac{1}{n^4k^5}
\end{align}
and $\lan e_n^2\ran-\lan e_n\ran^2\simeq 4/(\pi n)$. Thus, the
energy distribution has anticorrelations which  decrease with the
distance. The second scaling is clearly much more relevant to describe
the system, a fact that will become clearer in the following.

\subsection{The cumulative energy $E$}

It has been seen that the local structure of the stationary energy
field is of course affected by the injection mechanism. Nevertheless the preceding preliminary computations
highlighted a very important ``boundary layer'' phenomenon. To precise
this, let us define the \textit{cumulative energy} $E$:
\begin{align}
  E=\sum_{n=0}^{N-1}(1-\sig_n\sig_{n+1})
\end{align}
which is nothing but the energy of the $N$ first spins. The average of
$E$ and its fluctuations share the same following properties:
\begin{itemize}
  \item they are $O(\log N)$ when $N\rightarrow\infty$
\item their leading term is independent of the injection mechanism.
\end{itemize}
To see this, let us remind first that $\lan E\ran\sim\frac{2}{\pi}\log
N$ (from eq. \myref{moy_en}). As to the fluctuations, the
preceding section allows to write
\begin{align}
 \Delta E^2&\equiv\lan E^2\ran-\lan E\ran^2\\
&\sur{\sim}{N\rightarrow\infty}2\sum_{0\leq
  n<p<N}\left[-c_{n,p+1}+c_{n,p}+c_{n+1,p+1}-c_{n+1,p}+c_{n,p+1}c_{n+1,p}-c_{n,p}c_{n+1,p+1}\right]+\sum_{0\leq
  n<N}[\lan e_n^2\ran-\lan e_n\ran^2]\\
&\sur{\sim}{N\rightarrow\infty} \frac{8}{\pi}\left(1-\frac{2}{\pi}\right)\log N
\end{align}
The key point of this paper is to remark that these properties are
actually shared by all cumulants of the distribution of $E$. We do
not intend to give a demonstration for this, but owing to the
arguments presented in the preceding subsection, this conjecture is
reasonable: in an event of $2m$ walkers starting from sites located
asymptotically far from the boundary, the probability associated with
scenarios where one or several pairs of walkers reach the boundary
is either negligible or cancelled out by the cumulant structure of the mean. 

\bigskip

From the physical point of view, this property reflects a boundary
layer structure: the injection mechanism provides energy into the
system which spreads out quite efficiently despite the dissipation
mechanism; however the dissipation is strongly efficient near the
boundary and smoothes out the ``memory'' of the injection details. In
other words the dissipation sets up an autosimilarity
regime for the energy, which is independent of the amount transfered
to the system. 

From the analytical point of view, this property has two important
consequences. First, we shall see that there is a large deviation function
({\bf ldf}) $f$ associated with $E$, namely
\begin{align}
  \exists \ f, \ \lim_{N\rightarrow \infty}\frac{\log P(E/\log
  N=\zeta)}{\log N}=f(\zeta)
\end{align}
and secondly this ldf is \textit{independent} of the injection
mechanism, for the cumulants of $E$ are directly related to the Taylor
expansion coefficients of $f$. As already noticed
\cite{jeanlangevin,derridabodineau}, here again the ldf captures the
essential features of an observable and fades away some ``irrelevant''
details: the summation process tends to ``universalize'' the
different possible behaviours in some way. Moreover, this property
allows the analytical computation of $f(\zeta)$, which is the subject of
the subsequent section.

\section{The large deviation function of $E$}

As $f$ does not depend on the injection mechanism, we will consider
henceforth the limit $\la\rightarrow\infty$ (extremely quick flipping
of $\sig_0$). To get $f$, we will compute first the characteristic
function $G(\mu)=\lan \exp(-\mu E)\ran$, which is the Laplace
transform of the probability $P(E)$; we will show that $G(\mu)$ is
dominated by a term $\exp[g(\mu)\log N]$. Once $g$ is known, the
Laplace inversion formula shows that $f(\zeta)$ is given by a Legendre
transformation:
\begin{align}
  f(\zeta)=\max_{\mu}\left(\intvide g(\mu)+\mu \zeta\right)
\end{align}
provided that no analyticity breaking of the prefactors of the
exponentials in $G$ occurs at the saddle point in the $\mu$ space (see
\cite{jeanlangevin} for instance). We will assume in the following
that no such problem arises, which is the most often case.

\bigskip

Using a classical trick of Ising spin systems and defining $\tau=(\exp(-2\mu)-1)/2$, one can write $G$ as
\begin{align}
  G(\mu)&=\left\lan\prod_{i=0}^{N-1}e^{-\mu(1-\sig_i\sig_{i+1})}\right\ran=\left\lan\prod_{i=0}^{N-1}[1+\tau(1-\sig_i\sig_{i+1})]\right\ran\\
&=1+\tau\sum_{i}\lan
  1-\sig_i\sig_{i+1}\ran+\tau^2\sum_{i<j}\lan(1-\sig_i\sig_{i+1})(1-\sig_j\sig_{j+1})\ran+\tau^3\ldots\\
&\equiv 1+\tau S_1+\tau^2 S_2+\tau^3S_3+\ldots
\end{align}
In the limit $\la\rightarrow\infty$, all
the averages in the preceding sum can be interpreted as probabilities
related to walkers. For instance, $\lan 1-\sig_i\sig_{i+1}\ran$ is just
the probability $c_{i,i+1}$ that two walkers emerging from $i$ and
$i+1$ do not meet before reaching the site $0$; similarly
$\lan(1-\sig_i\sig_{i+1})(1-\sig_j\sig_{j+1})\ran$ is (for $i\neq j$) the probability
that during the wandering of four walkers emerging from sites labeled
$i$, $i+1$, $j$, $j+1$, walkers $i$ and $i+1$ from one side \textit{and}
walkers $j$ and $j+1$ from the other side have not collapsed (this
interpretation is also valid if $i+1=j$). We will denote this
probability $p_{I,J}$ in the following. We will see that the
probabilities $p_{I,J,K,\ldots}$ and therefore the terms $S_i$ can be
expressed in terms of the $c_{i,j}$ only.

\subsection{The first terms}

The term $S_1$ is simply $S_1=\sum_i c_{i,i+1}$, where the summation
ranges from $i=0$ to $i=N-1$.
The term $S_2$ can also be expressed in terms of the $c_{i,j}$. To
this end, let us consider $p_{I,J}$. The event $(I,J)$, which
is ``walkers $i$ and $i+1$ do not collapse, as well as walkers $j$ and
$j+1$'' is equivalent to the event ``$(i,i+1,j,j+1)$ or
$(i,(i+1j),j+1$'', such that (remind that a comma holds for ``have not
met'' whereas an absence of comma for ``have met'')
\begin{align}
  p_{I,J}=c_{i,i\+1,j,j+1}+c_{i,i\+1\mspace{1mu}j,j+1}
\end{align}
This probability can be transformed slightly. We have the important
relations (actually generalizable to any situation)
\begin{align}
\begin{split}
  c_{i,jk,\ell}+c_{i,j,k\ell}+c_{i,j,k,\ell}&=c_{i,j,\ell}\\
c_{i,j,k\ell}+c_{i,j,k,\ell}&=c_{i,j,k}\end{split}\label{actually}
\end{align}
which are valid whatever $i\leq j\leq k\leq\ell$ (essentially they
express the fact that if one enumerates correctly what can happen to
an emerging path ---all remaining paths being besides fixed---, and
sum the corresponding probabilities, one gets the probability of the
event whence that path is eliminated). They show how to
express $c_{\ldots}$ where some indices ``stick'' together, in terms
of $c_{\ldots}$ with no sticking. 

This yields for $p_{I,J}$ and $S_2$:
\begin{align}
  p_{I,J}&=c_{i,i+1,j,j+1}+c_{i,i+1,j+1}-c_{i,i+1,j}\\
S_2&=\sum_{i<j}c_{i,i+1,j,j+1}+\sum_i c_{i,i+1,N}
\end{align}
The complexity of the terms increases very fast with the number of
indices:
\begin{align}
  p_{I,J,K}&=c_{i,i+1,j,j+1,k,k+1}+c_{i,i+1j,j+1,k,k+1}+c_{i,i+1,j,j+1k,k+1}+c_{i,i+1j,j+1k,k+1}\\
S_3&=\sum_{i<j}c_{i,i+1,j,j+1}+\sum_{i<j<k}c_{i,i+1,j,j+1,k,k+1}+\sum_{i<j}c_{i,i+1,j,j+1,N}
\end{align}
(the sum $S_3$ is obtained after a quite lengthy computation!). The
next order requires some patience, and one gets
\begin{align}
  S_4&=\sum_{i<j<k<\ell}c_{i,i+1,j,j+1,k,k+1,\ell,\ell+1}+2\sum_{i<j<k}c_{i,i+1,j,j+1,k,k+1}+\sum_{i<j<k}c_{i,i+1,j,j+1,k,k+1,N}+\sum_{i<j<k}c_{i,i+1,j,j+1,N}
\end{align}
No evident regularity emerges from these first terms. It is worth
noting that a rewriting of these terms is possible, since it is
possible to write a term $c_{\ldots}$ with odd number of indices as a
sum 
of $c_{\ldots}$ with even number of indices, but this rewriting makes the
situation all but more transparent. 

\subsection{Resummation of $G$}

 A resummation of $G(\mu)$
is nevertheless possible. To do this, it is convenient to
introduce a schematic representation of the terms appearing in the
$S_j$. With evident definitions, we introduce the notations $C^{(j)}$
and $C_N^{(j)}$:
\begin{align}
  S_1&=C^{(1)}\\
S_2&=C^{(2)}+C_N^{(1)}\\
S_3&=C^{(3)}+C^{(2)}+C_N^{(2)}\\
S_4&=C^{(4)}+2C^{(3)}+C_N^{(3)}+C_N^{(2)}
\end{align}
We will show that a recursion relation exists between $S_{n+2}$,
$S_{n+1}$ and $S_n$. The starting point is, besides the remark that
$S_n$ is a sum of terms $C^{(j)}$ and $C^{(j)}_N$,  the following relation,
obtained along similar lines as eq. \myref{actually}:
\begin{align}
  p_{I,J,K}&=p_{I,J\text{ and }j+1,k,k+1}+p_{I,J\text{ and
  }j+1k,k+1}\\
&=p_{I,J\text{ and }j+1,k+1}-p_{I,J\text{ and }j+1,k}+p_{I,J\text{ and }j+1,k,k+1}
\end{align}
In this relation, $p_{I,J\text{ and }j+1,k}$ terms the probability
that $i$ and $i+1$, $j$ and $j+1$, $j+1$ and $k$ do not meet pair by
pair. The other terms have similar definitions. Moreover, we
particularize the recursion relation on $p_{I,J,K}$, but it is evident
that  a such relation is also true if some capital indices are
implied at the left of $I$: this relation is true at any order
actually. 
As a result,
\begin{align}
  \sum_{i<j<k}p_{I,J,K}&=\sum_{i<j}p_{I,J\text{ and
  }j+1,N}+\sum_{i<j<k}p_{I,J\text{ and }j+1,k,k+1}
\end{align}
Naively, we are tempted to deduce from this relation the formal law to
pass from $S_n$ to $S_{n+1}$: ``put an extra $N$ on one side,
increment the rank on an other, and add''. This rule is in fact not
simple to implement for a recursion, since one does not know a priori how to interpret
terms like $(C^{(n)}_{N})_N$ and $(C^{(n)}_N)_{n\rightarrow n+1}$. Let
us see how to do this. Again, we examplify the demonstration on a
certain rank, but it is obviously valid for any.

We have 
\begin{align}
  S_2&=\sum_{i<j}p_{I,J}\\
S_3&=\sum_{i<j<k}\left[p_{I,J\text{ and }j+1,k+1}-p_{I,J\text{ and }j+1,k}\right]+p_{I,J\text{ and }j+1,k,k+1}\label{s3}
\end{align}
So, the passage from $S_n$ to $S_{n+1}$ is obtained by the formal
linear operator: $\mathcal{L}=(\ldots)_{k+1}-(\ldots)_k+(\ldots)_{k,k+1}$. Let us
write now $S_2$ in terms of $C^{(2)}$ and $C^{(1)}_N$:
\begin{align}\label{eq2}
  S_2&=\sum_{i<j}\left[p_{I\text{ and }i+1,j+1}-p_{I\text{ and }i+1,j}\right]+\sum_{i<j}p_{I\text{ and }i+1,j,j+1}
\end{align}
The question now is: can we apply the operator $\mathcal{L}$ term by
term to the equation \myref{eq2}~? We have
\begin{align}
  \mcal{L}\left[p_{I\text{ and }i+1,j+1}\right]&=p_{I\text{ and
  }i+1,j+1,k+1}-p_{I\text{ and }i+1,j+1,k}+p_{I\text{ and
  }i+1,j+1,k,k+1}\\
\mcal{L}\left[p_{I\text{ and }i+1,j}\right]&=p_{I\text{ and
  }i+1,j,k+1}-p_{I\text{ and }i+1,j,k}+p_{I\text{ and
  }i+1,j,k,k+1}\\
\mcal{L}\left[p_{I\text{ and }i+1,j,j+1}\right]&=p_{I\text{ and
  }i+1,j,j+1,k+1}-p_{I\text{ and }i+1,j,j+1,k}+p_{I\text{ and
  }i+1,j,j+1,k,k+1}
\end{align}
On the other hand, the generic term of $S_3$ (cf. eq \myref{s3}) can
be expanded ``inside'' using the relations:
\begin{align}
  p_{I,J\text{ and }j+1,k+1}&=p_{I\text{ and }i+1,j,j+1,k+1}+p_{I\text{
  and }i+1j,j+1,k+1}\\
&=p_{I\text{ and }i+1,j,j+1}-p_{I\text{ and }i+1,j,k+1}+p_{I\text{ and
  }i+1,j+1,k+1}\\
p_{I,J\text{ and }j+1,k}&=p_{I\text{ and }i+1,j,j+1}-p_{I\text{ and }i+1,j,k}+p_{I\text{ and
  }i+1,j+1,k}\\
p_{I,J\text{ and }j+1,k,k+1}&=p_{I\text{ and
  }i+1,j,j+1,k+1}-p_{I\text{ and }i+1,j,j+1,k}+p_{I\text{ and
  }i+1,j+1,k,k+1}\nonumber\\
&\ \ \ \quad -p_{I\text{ and }i+1,j,k,k+1}+p_{I\text{ and }i+1,j,j+1,k,k+1}
\end{align}
A rapid inspection shows that  application of $\mcal{L}$ term
by term in \myref{eq2} gives the correct result. Moreover, one sees
that $\mcal{L}[C^{(1)}_N]=C^{(2)}$ and $\mcal{L}[C^{(2)}]=C^{(3)}+C^{(2)}_N$
and similar relations hold at any order:
\begin{align}
  \mcal{L}\left[C^{(n)}_N\right]&=C^{(n+1)}\\
\mcal{L}\left[C^{(n)}\right]&=C^{(n+1)}+C^{(n)}_N
\end{align}
A couple of polynomials $(P_n,Q_n)$ can be associated to $S_n$: $P_n$
represents the $C$ terms and $Q_n$ the $C_N$ terms, with the mapping
$C^{(m)}\leftrightarrow X^m$. Thus,
\begin{align}
(P_1,Q_1)&=(X,0)\\
(P_2,Q_2)&=(X^2,X)\\
(P_3,Q_3)&=(X^3+X^2,X^2)\\
\ldots\\
(P_{n+1},Q_{n+1})&=(XP_n+XQ_n,P_n)
\end{align}
whence $P_{n+1}=XP_n+XP_{n-1}$. Symbolically, this recursion can be
solved for $P_n$, owing to $P_1=X,P_2=X^2$:
\begin{align}
  P_n&=\frac{X}{\sqrt{X^2+4X}}\left[\left(\frac{X+\sqrt{X^2+4X}}{2}\right)^n-\left(\frac{X-\sqrt{X^2+4X}}{2}\right)^n\right]
\end{align}
$Q_n$ is simply deduced from the formula $Q_n=P_{n-1}$. Symbolically,
it yields
\begin{align}
  \sum_{n=1}^\infty P_n\tau^n&=\frac{X\tau}{1-X(\tau+\tau^2)}\\
&=\frac{1}{1+\tau}\sum_{n=1}^\infty(\tau+\tau^2)^nX^n
\end{align}
and
\begin{align}
  \sum_{n=1}^\infty Q_n\tau^n&=\frac{\tau}{1+\tau}\sum_{n=1}^\infty(\tau+\tau^2)^nX^n
\end{align}
From these results is derived a rewriting of $G(\mu)$:
\begin{align}
  G(\mu)=&1+\frac{1}{1+\tau}\sum_{n=1}^\infty(\tau+\tau^2)^nC^{(n)}+\frac{\tau}{1+\tau}\sum_{n=1}^\infty(\tau+\tau^2)^nC_N^{(n)}\\
=&1+\frac{1}{\tau+1}\left[(\tau+\tau^2)\sum_i
  c_{i,i+1}+(\tau+\tau^2)^2\sum_{i<j}c_{i,i+1,j,j+1}+\ldots\right]\nonumber\\
&\  +\frac{\tau}{\tau+1}\left[(\tau+\tau^2)\sum_ic_{i,i+1,N}+(\tau+\tau^2)^2\sum_{i<j}c_{i,i+1,j,j+1,N}+\ldots\right]
\end{align}
The second sum can be slightly transformed : it can be shown (see
Appendix) that
\begin{align}
  c_{i_1,i_1+1,i_2,i_2+1,\ldots,i_n,i_n+1,N}=&c_{i_1,i_1+1,i_2,i_2+1,\ldots,i_n,i_n+1}-c_{i_1,i_1+1,i_2,i_2+1,\ldots,i_n,N}+\ldots\nonumber\\
&\
  -c_{i_1,i_2,i_2+1,\ldots,i_n+1,N}+c_{i_1+1,i_2,i_2+1,\ldots,i_n,i_n+1,N}
\end{align}
whence
\begin{align}
C^{(n)}_N=&\sum_{0\leq i_1<i_2<\ldots<i_n}c_{i_1,i_1+1,i_2,i_2+1,\ldots,i_n,i_n+1}-\sum_{0<i_2<\ldots<i_n}c_{0,i_2,i_2+1,\ldots,i_n,i_n+1,N}\nonumber\\
=&C^{(n)}-C_N^{(n-1)}+\sum_{i_3<\ldots<i_n}c_{1,i_3,\ldots,i_n,i_n+1,N}\label{oddtoeven}
\end{align}
(otherwise explicitely stated, let us recall that the sums start at
site $0$). Whence
\begin{align}
  \sum_{n=1}^\infty(\tau+\tau^2)^nC_N^{(n)}&=-\frac{\tau+\tau^2}{1+\tau+\tau^2}+\frac{1}{1+\tau+\tau^2}\sum_{n\geq1}(\tau+\tau^2)^nC^{(n)}+\frac{(\tau+\tau^2)^2}{1+\tau+\tau^2}\sum_{n\geq0}(\tau+\tau^2)^nC_{1(n)N}
\end{align}
where
\begin{align}\label{c1pn}
  C_{1(n)N}&=\sum_{0<i_1<i_2<\ldots<i_n}c_{1,i_1,i_1+1,i_2,i_2+1,\ldots,i_n,i_n+1,N}
\end{align}
As a result, one has
\begin{align}
  G(\mu)&=\frac{1+\tau}{1+\tau+\tau^2}\left[1+\sum_{n\geq
  1}(\tau+\tau^2)^nC^{(n)}+\tau^3\sum_{n\geq 0}(\tau+\tau^2)^nC_{1(n)N}\right]
\end{align}

This rewriting is  fruitful, since the sums are directly
related to the Pfaffian theory: as explained in the Appendix, we can
write
\begin{align}
  G(\mu)&=\frac{1+\tau}{1+\tau+\tau^2}\sqrt{\det[1+A(\eta,\eta')C]}\ \
  \text{with}\\
\eta&=-\tau-\tau^2\\
\eta'&=\frac{\tau^2}{1+\tau}
\end{align}
(the definitions of $A$ and $C$ are given in the Appendix).

\subsection{Asymptotic behaviour of $G$}

In \cite{derridahakimpasquier}, the authors showed that the leading term of
$\log\det[1+A(\eta,\eta')C]$ remains unchanged if we put $\eta'=0$. In
fact, in their case, a difficulty arises, due to the fact that the subdominant
term (where $\eta'$ appears) diverges in a certain range of parameters and provides an analytic
continuation of the persistence exponent. Here, the computation of the
subdominant term cannot be performed exactly because the matrix $AC$
does not have an exact continuous limit, but  we think that it is
precisely this fact that
prevents the subdominant term to interfere in the analytic properties
of the ldf. Anyway, we did not find any analytical problem in the ldf
(see below),
and interpreted this as a ``gentle'' behaviour of the subdominant
term. In the following the replacement $\eta'=0$ is implicitely made.

\bigskip

Using $\log\det(M)=\text{Tr}\log(M)$, we get
\begin{align}
  \log G(\mu)&\sim-\demi\sum_{p=1}^\infty\frac{(-1)^p}{p}\text{Tr}\left[(A(\eta,0)C)^p\right]
\end{align}
From the expressions of $A$ and $C$, we have:
\begin{align}
  \text{Tr}\left[(A(\eta,0)C)^p\right]\sur{\sim}{N\rightarrow\infty}
  \left(\frac{8}{\pi}(\tau+\tau^2)\right)^p\sum_{1\leq
  i_1,\ldots,i_p\leq N}\frac{i_1i_2\ldots i_p}{(i_1^2+i_2^2)(i_2^2+i_3^2)\ldots(i_p^2+i_1^2)}
\end{align}
We get this first expression by assuming that the leading term is
obtained when all indices are large (cf. \cite{derridahakimpasquier}
and see below). We
assume temporarily that this term is $O(\log N)$; thus we have necessarily for any $a>1$
\begin{align}
  \sum_{i_1,\ldots,i_p\leq N}\frac{i_1i_2\ldots
  i_p}{(i_1^2+i_2^2)(i_2^2+i_3^2)\ldots(i_p^2+i_1^2)}&\sim\frac{\log
  N}{\log a}\sum_{N\leq i_1,\ldots,i_p\leq Na}\frac{i_1i_2\ldots
  i_p}{(i_1^2+i_2^2)(i_2^2+i_3^2)\ldots(i_p^2+i_1^2)}\\
&\sim \frac{\log
  N}{\log a}\int_0^{\log a}  \frac{du_1\ldots du_p}{(1+e^{2(u_2-u_1)})(1+e^{2(u_3-u_2)})\ldots(1+e^{2(u_1-u_p)})}
\end{align}
Let us now consider the large $a$ limit in the integral and perform
the change of variable $v_1=u_1$, $v_2=u_2-u_1$, \ldots
$v_p=u_p-u_{p-1}$ (its Jacobian is $1$). Defining
$s(x)=e^{-x}/(1+e^{-2x})=1/2\cosh(x)$, we have
\begin{align}
  \int_0^{\log a}  \frac{du_1\ldots
  du_p}{(1+e^{2(u_2-u_1)})\ldots(1+e^{2(u_1-u_p)})}&=\int_{-\infty}^\infty\!\!
  dy \int_0^{\log a}\!\! dv_1 \int dv_2\ldots dv_p \
  \de(v_2\+\ldots\+v_p\!-\!y)s(v_2)\ldots s(v_p)s(y)\\
&\sur{\sim}{a\rightarrow\infty}\log a \times \int_{-\infty}^\infty
  \frac{dk}{2\pi}\left[\int_{-\infty}^\infty dv e^{ikv}s(v)\right]^p\label{pouet}
\end{align}
Before proceeding further, let us show that this term can be expressed
differently, which will make the $\log N$ dependence explicit:
\begin{align}
\sum_{i_1,\ldots,i_p\leq N}  \frac{i_1i_2\ldots
  i_p}{(i_1^2+i_2^2)(i_2^2+i_3^2)\ldots(i_p^2+i_1^2)}&\sim p\sum_{i_p=1}^{N}\frac{1}{i_p^p}\sum_{i_1,\ldots,i_{p-1}\leq i_p}\frac{(i_1/i_p)\ldots
  (i_{p-1}/i_p)}{((i_1/i_p)^2+(i_2/i_p)^2)\ldots((i_1/i_p)^2+1)}\\
&\sim p\log(N)\int_0^1dx_1\ldots dx_{p-1}\frac{x_1\ldots x_{p-1}}{(x_1^2+x_2^2)\ldots(x_{p-1}^2+1)(1+x_1^2)}
\end{align}
(because the general term of the $i_p$ sum is asymptotically
equivalent to the integral times
$1/i_p$). This expression is however not as useful as the preceding, for it
leads to an expression involving a solution of a Wiener-Hopf problem.

\medskip

\begin{figure}[h]\label{ldffig}
  \begin{center}
    \resizebox{7cm}{!}{\includegraphics{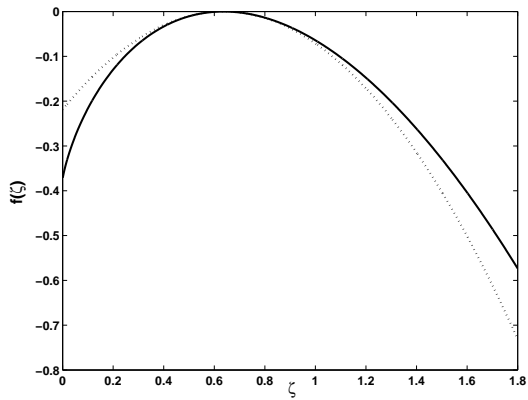}}
  \end{center}\caption{The large deviation function $f(\zeta)$ for the
    Ising model (solid
    line) and the best parabolic approximation
    $-(\zeta-2/\pi)^2/(2g''(0))$ (dots). The ldf does not continue in the range $\zeta<0$.}
\end{figure}

Coming back to $G(\mu)$, eq. \myref{pouet} and \cite{derridahakimpasquier} yield
\begin{align}
  \log G(\mu)&\sim \log(N)\times \int_{-\infty}^\infty \frac{dk}{4\pi}
  \log\left[1+\frac{8}{\pi}(\tau+\tau^2)\int_{-\infty}^\infty dv \
  e^{ikv}s(v)\right]\label{prolan}\\
\Rightarrow g(\mu)&= \frac{1}{8}-\left[\frac{\sqrt{2}}{\pi}\text{Arccos}\left(\frac{e^{-2\mu}}{\sqrt{2}}\right)\right]^2
\end{align}
 We verify that $g(0)=0$,  $-g'(0)=\lan E/\log N\ran=2/\pi$ and
$g''(0)=\frac{8}{\pi^2}(\pi-2)=(\lan E^2\ran-\lan E\ran^2)/\log N$,
 as it must be. The third derivative of $g$ is $64(3-\pi)/\pi^2$ on
 zero. In the figure \ref{ldffig}, the aspect of $f(\zeta)$ is shown
 and its shape is compared with a parabola. The fluctuations above the
 mean value are comparatively more probable than those below the mean
 value: intuitively a positive fluctuation of $E$ regresses by
 annihilation pair by pair of the (excess) domain walls, and can be
 sustained as soon as the domain walls  avoid
 each other in their erratic motion; on the contrary, a important negative fluctuation
 (i.e. $E<\lan E\ran$) will recede by energy refilling from the boundary, a
 process which is unavoidable if the population of domain walls is
 substantially lowered. This could explain qualitatively the
 counterclockwise tilt of the curve.

As regards the right asymptotics of $f(\zeta)$, it is easily
established (with the formula \myref{prolan} which is the analytic
continuation of $g(\mu)$) that $f(\zeta)\sim -(\pi\zeta/8)^2$; this
explains the vaguely parabolic look of $f$.

Finally let us notice that the $f(\zeta)$ stops abruptly at $\zeta=0$;
this means that the probability of observing an arbitrary small energy
$E$ decreases like $N^{-3/8}$ (up to a $1/\sqrt{\log N}$ term); it
illustrates the fact that an important ``clearance'' of the system due
to an efficient (in dissipation) fluctuation is not particularly unlikely. In  \cite{derridadoucotroche}
a same kind of ldf with a non diverging left branch was found  in a very different context.

\subsection{Generalization to (an)isotropic  Potts models}

The above calculations can be extended without difficulty to a $q$-states Potts
model (driven by a ``voter-model'' dynamics: in $[t,t+dt]$, each spin imitates its
left/right neighbour with a probability $dt$ and does nothing with
probability $1-2dt$) with an arbitrary input at $\sig_0$ (the only requirements
concerning this input is that it must be statistically stationary and
have a finite time scale; we call this arbitrariness ``anisotropic''
for the different colours of the Potts model are not necessarily equiprobable). The
energy is now defined as $e_j=2(1-\de_{\sig_j,\sig_{j+1}})$ (we put the
factor two to make the Ising model equivalent to the $q=2$ Potts). If
$P_0(\sig_0)$ is the stationary probability of the spin $\sig_0$, let
us define
\begin{align}
  b=2\times\left(1-\sum_{\ell=1}^q P_0(\ell)^2\right)
\end{align}
($b=1$ in the Ising model; $b=2$ corresponds to a $A+A\rightarrow A$
model). 

Actually, this generalization does not modify deeply the preceding
analysis : the reasoning on the coalescing walkers is unaffected by
the multiplication of the ``colours'', since either the walkers collapse,
which implies automatically that they ``carried'' the same colour
(remind that the walkers cannot be considered as domain walls or vice-versa), or
they do not collapse, and in that case their colours are statistically determined by
the $\sig_0$ process, which can be considered as before as an
ultrafast process.

It is easily shown that $\lan
E\ran\sim \log(N)2b/\pi$ and similarly the function $G$ is obtained
from the preceding by making the replacement $\tau\rightarrow b\tau$:
\begin{align}
  g(\mu)&= \frac{1}{8}-\left[\frac{\sqrt{2}}{\pi}\text{Arccos}\left(\frac{be^{-2\mu}+1-b}{\sqrt{2}}\right)\right]^2
\end{align}
\begin{figure}[h]\label{ldfb}
  \centerline{\resizebox{9cm}{!}{\includegraphics{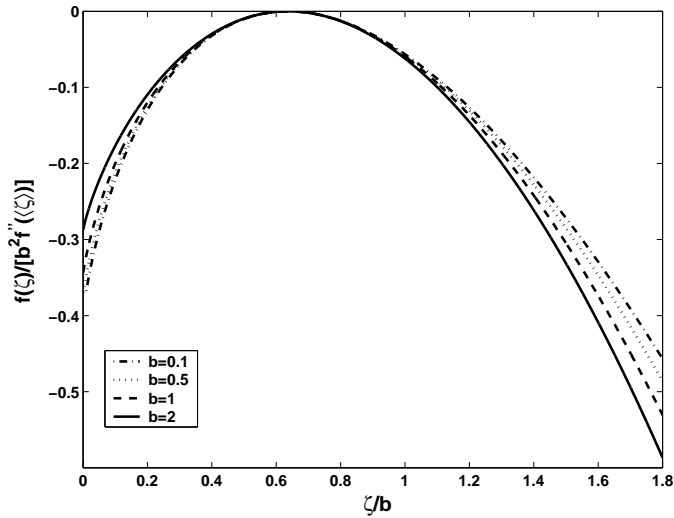}}}
\caption{Normalized ldf for different values of $b$ for the
  anisotropic Potts models: The plots are $f(\zeta)/[b^2f''(\lan
  \zeta\ran)]$ as a function of $\zeta/b$. The Ising model corresponds
  to $b=1$.}
\end{figure}
In the figure \ref{ldfb} were plotted the ldf for several values of
$b$. To compare these functions we normalized the coordinates, so that
their maximum is located at $2/\pi$ and their curvature at the
maximum is $-1$. Once normalized,  these functions look very similar,
and the physical informations are mainly contained in the first two
moments: we have $\lan E\ran/\log N\sim 2b/\pi$ and
$(\lan E^2\ran-\lan E\ran^2)/\log N\sim 4b[\pi+(\pi-4)b]/\pi^2$. As a
result, irrespective of the details of the injection, the
anisotropic Potts models verify asymptotically the simple relation
\begin{align}
  \frac{\De E^2}{\lan E\ran}=\left[\intvidepetit (\pi-4)\frac{\lan
  E\ran}{\log N}+2\right]
\end{align}
A natural question would be to know if such a relation can still be verified
if intrinsic dynamics, although dissipative, is not
equivalent to a voter model.

\section{Nonequilibrium ``free energy''}

As it was possible to compute the large deviation function of $E$, we
can ask whether  a more precise description
of the fluctuations of the energy field could be tractable. The first idea is to form
$n_0$ groups of $N$ consecutive spins and try to compute the
probability to observe a certain coarse-grained
energy profile $(E_1,\ldots,E_{n_0})$ when $N$ is large. But this idea does not take
into account the great intrinsic inhomogeneity of the dissipative
systems: if the typical energy of the $N$ first spins is $O(\log N)$,
the typical energy of the spins from indices $N$ to $2N$ is only
$O(1)$, which prevents to give a satisfactory description in terms of
large deviation function. The second idea is to rectify by hand this
inhomogeneity, considering for instance variables like $\sum_i
ie_i$. But it is easy to show that they do not obey a large deviation
theorem in the large $N$ limit. Actually the only way to construct a
large deviation function capable of describing more precisely the energy
fluctuations is to consider \textit{unequal} blocks of spins: the
first group gather spins $\sig_0$ to $\sig_{N-1}$, the second spins
from $\sig_{N}$ to $\sig_{N^2-1}$,\ldots, the last, spins from
$\sig_{N^{n_0-1}}$ to $\sig_{N^{n_0}-1}$. We are going to show that
this inhomogeneous grouping of spins allows a relevant description in
terms of ldf.

\bigskip

For the sake of simplicity, we come back to the Ising model, but the
generalization of the preceding paragraph could be processed along similar
simple lines. As well, we forget for a moment the particular way to
group the spins together and consider something more general:
\begin{align}
  H(\pmb{\mu})&=\left\lan\exp\left(-\sum_{i=0}^{M-1}\mu_i(1-\sig_i\sig_{i+1})\right)\right\ran
\end{align}
$H$ is a characteristic function for $M$ spins where $\mu$ is replaced by an arbitrary
``field'' $\pmb{\mu}$. A multidimensional Laplace inversion performed
over $\pmb{\mu}$ would give the probability of the energy density field. 

\bigskip

A similar procedure as for $G$ leads to
\begin{align}
  H(\pmb{\mu})&=1+\sum_i\tau_ip_I+\sum_{i<j}\tau_i\tau_jp_{I,J}+\ldots\\
&\equiv 1+T_1+T_2+\ldots
\end{align}
 with $\tau_j=(e^{-2\mu_j}-1)/2$. 
The resummation procedure is more complicated than for $G$. Let us
define $\Delta_j=\tau_{j-1}-\tau_j$. The expansion of $T_2$ using the
recursion relation \myref{recrel} leads to
\begin{align}
  T_2&=\sum_{i<j}\tau_i\Delta_jp_{I\text{ and }i+1,j}+\sum_{i<j}\tau_i\tau_jp_{I\text{ and }i+1,j,j+1}
\end{align}
Symbolically, the ``$\tau$'' term of $T_1$ is transformed to a
``$\tau\Delta+\tau\tau$'' term at the next step. A lengthy computation
shows that 
\begin{align}
  T_3&=\sum_{i<j<k}\tau_i\Delta_j\Delta_kp_{I\text{ and
  }i+1,j,k}+\sum_{i<j<k}\tau_i\Delta_j\tau_kp_{I\text{ and
  }i+1,j,k,k+1}+\sum_{i<k}\tau_i\tau_k^2p_{I\text{ and
  }i+1,k,k+1}\nonumber\\
&\phantom{aaaaaa}\ \ +\sum_{i<j<k}\tau_i\tau_j\Delta_kp_{I\text{ and
  }i+1,j,j+1,k}+\sum_{i<j<k}\tau_i\tau_j\tau_kp_{I\text{ and
  }i+1,j,j+1,k,k+1}\\
&=''\wtau\wDe\wDe+\wtau\wDe\wtau+\wtau\wtau_2+\wtau\wtau\wDe+\wtau\wtau\wtau''
\end{align}
where the last line is a symbolic rewriting of $T_3$. Besides, the computation
shows  that $(\ldots)\wDelta$ is transformed into
$(\ldots)(\wDelta\wDelta+\wDelta\wtau+\wtau_2)$. Thus, as for $G$, these
simple rules allow us to generate any order, which are rapidly
increasing in complexity:
\begin{align}
  T_3&=\wtau\wDe\wtau+\wtau\wDe\wDe+\wtau\wtau_2+\wtau\wtau\wDe+\wtau\wtau\wtau\\
\Rightarrow T_4&=\wtau\wDe\wtau\wDe+\wtau\wDe\wtau\wtau+\wtau\wDe\wDe\wDe+\wtau\wDe\wDe\wtau+\wtau\wDe\wtau_2+\wtau\wtau_2\wDe+\wtau\wtau_2\wtau+\wtau\wtau\wDe\wtau+\wtau\wtau\wDe\wDe+\wtau\wtau\wtau_2+\wtau\wtau\wtau\wDe+\wtau\wtau\wtau\wtau
\end{align}
These terms can be grouped into four categories: there are terms
(called ``$u_n$'' in the following) without any $\wDe$, terms with a
single and terminal $\wDe$ (called ``$v_n$''), terms like $\wtau\wDe\wDe\wtau$ with multiple
internals $\wDe$ but not ending with a $\wDe$ (called ``$w_n$''), and
terms with multiple $\wDe$ ending with a $\wDe$ (called
``$z_n$''). Recursions relations exist between these fours sequences:
one has for $u$ and $v$:
\begin{gather}
  u_1=\wtau,\ \ \ 
u_2=\wtau\wtau,\ \ \ 
u_{n+1}=u_n\wtau+u_{n-1}\wtau_2\\
v_1=0,\ \ \ 
v_{n+1}=u_n\wDelta
\end{gather}
Let us recall that 
\begin{itemize}
  \item the elements $\wtau$, $\wtau_2$ and $\wDe$ of the symbolic algebra
  do not commute.
\item the element $\wtau_2$ is not equivalent to $\wtau\wtau$
\end{itemize}
Formally, one can sum up the $u_n$ and $v_n$ terms:
\begin{align}
  U\equiv \sum_{n=1}^\infty u_n&=\wtau(1-\wtau-\wtau_2)^{-1}\\
V\equiv\sum_{n=1}^\infty v_n&=\wtau(1-\wtau-\wtau_2)^{-1}\wDe
\end{align}
The $w$ and $z$ terms obey also recursion relations:
\begin{align}
  w_{n+1}&=v_n\wtau+z_n\wtau+w_n\wtau+z_n\!\!\not\!\!\wDe\wtau_2\\
z_{n+1}&=(w_n+z_n+v_n)\wDe
\end{align}
where $z_n\!\!\not\!\!\!\wDe$ means ``$z_n$ without its final $\wDe$''. Defining
$W=\sum_{n\geq 3}w_n$ and $Z=\sum_{n\geq3}z_n$, a simple computation
gives
\begin{align}
  H(\pmb{\mu})&=1+U+V+W+Z\\
&=1+\wtau\left(1-\wtau-\wtau_2-\wDe\right)^{-1}
\end{align}
This simple formal result hides a real complexity: let us recall that
for instance
\begin{align}
  \wtau\wDelta\wtau_2\wDelta&=\sum_{i<j<k<\ell}\tau_i\De_j\tau^2_k\De_\ell
  c_{i,i+1,j,k,k+1,\ell}
\end{align}
A useful rewriting of $H$ is
\begin{align}
  H(\pmb{\mu})&=1+\wtau\sum_{p=0}^\infty\left[(1-\wtau-\wtau_2)^{-1}\wDe\right]^p(1-\wtau-\wtau_2)^{-1}
\end{align}

\subsection{Recovering $G$}
If all the $\mu_j$ are constant, this sum is considerably simplified:
in fact, all $\De_j$ but $\De_N=\tau$ are zero, and $\wtau_2=\tau\wtau$ such that
\begin{align}
  H(\mu)&=1+\wtau(1-\wtau-\wtau_2)^{-1}+\wtau(1-\wtau-\wtau_2)^{-1}\wDe\\
&=1+\tau\sum_{p=0}^\infty(\tau+\tau^2)^pC^{(p+1)}+\tau^2\sum_{p=0}^\infty
  (\tau+\tau^2)^pC^{(p+1)}_N=G(\mu)
\end{align}
$G(\mu)$ is recovered. 

\subsection{Two groups of spins}

Let us now consider a situation with two
clusters of $\mu_j$: $\mu_j=\mu$ if $j\in[0,N-1]$ and $\mu_j=\mu'$ if
$j\in[N,N^2-1]$. In that case, two $\De_j$ only are nonzero
($\De_N=\tau-\tau'$ and $\De_{N^2}=\tau'$) and $H$ is expressed as
\begin{align}
   H(\mu,\mu')&=1+\wtau(1-\wtau-\wtau_2)^{-1}+\wtau(1-\wtau-\wtau_2)^{-1}\wDe(1-\wtau-\wtau_2)^{-1}+\wtau(1-\wtau-\wtau_2)^{-1}\wDe(1-\wtau-\wtau_2)^{-1}\wDe
\end{align}
We can write $H(\mu,\mu')$ as $H(\mu,\mu')=H(0,\mu')+\De H$. It is
easy to see that  $H(0,\mu')$ (resp. $\De H$) is made with all
terms where the first $\tau_i$ of the sums is $\tau'$
(resp. $\tau$). Defining
\begin{align}
  \widetilde{H}(\mu,\mu')&\equiv
  1+(\wtau+\wtau_2)(1-\wtau-\wtau_2)^{-1}+(\wtau+\wtau_2)(1-\wtau-\wtau_2)^{-1}\wDe(1-\wtau-\wtau_2)^{-1}\nonumber\\
&\ \ \ +(\wtau+\wtau_2)(1-\wtau-\wtau_2)^{-1}\wDe(1-\wtau-\wtau_2)^{-1}\wDe
\end{align}
we get
\begin{align}
  \widetilde{H}&=1+(1+\tau)\De H+(1+\tau')[H(0,\mu')-1]\\
&=-\tau'+(1+\tau)H+(\tau'-\tau)H(0,\mu')
\end{align}
$\widetilde{H}$ is interesting for it is easier to compute than $H$;
the last expression shows that once $\widetilde{H}$ is computed, so
is $H$, since $H(0,\mu')$ can be estimated along lines similar to
those we followed for $G$.

\bigskip

Let us define $\xi=\tau+\tau^2$ and $\xi'=\tau'+{\tau'}^2$. Similarly
$\wxi$ (resp. $\wxi'$) is the
operator
$\wtau+\wtau_2$ applied on indices less (resp. greater) than $N$.  As an example,
\begin{align}
  \wxi^2\wxi'=\xi^2\xi'\sum_{0\leq i_1<i_2<N\leq j_1}c_{i_1i_1\+1,i_2,i_2\+1,j_1,j_1\+1}
\end{align}
Similarly $\wDe_N$ and $\wDe_{N^2}$ term the $\wDe$ operator where the
index of application is precised. We have 
\begin{align}
\widetilde{H}(\mu,\mu')&=1+\sum_{p+q> 0}\wxi^p{\widetilde{\xi}'\null^q}+\sum_{p\geq 1,q\geq 0}\wxi^p\wDe_N\wxi'\null^{q}+\sum_{p+q> 0}\wxi^p\wxi'\null^{q}\wDe_{N^2}+\sum_{p\geq 1,q\geq 0}\wxi^p\wDe_N\wxi'\null^{q}\wDe_{N^2}
\end{align}
Terms with a single $\wDe$ can be further simplified using \myref{oddtoeven}:
\begin{align}
  \wxi^p\wDe_N{\wxi'}\null^q&=\xi^p{\xi'}^q\De_N\left[-C_{(p),N,(q-1),N^2}+C_{(p),(q)}-C_{(p-1),N,(q)}+C_{1,(p-2),N,(q)}\right]\\
\wxi^p{\wxi'}\null^q\wDe_{N^2}&=\xi^p{\xi'}^q\De_{N^2}\left[C_{(p),(q)}-C_{(p),N,(q-1),N^2}+C_{(p-1),N,(q),N^2}-C_{(p-1),(q),N^2}+C_{1,(p-2),(q),N^2}\right]
\end{align}
The definition of the symbols $C_{\ldots}$ is generalized from
\myref{c1pn}; $C_{(p),(q)}$ deserves however a precision: it
holds for a summation over $0\leq
i_1<\ldots<i_p<N\leq j_1<\ldots<j_q<N^2$ (instead of
$\ldots<N<j_1<\ldots$). After a computation, we get
\begin{align}
  \widetilde{H}(\mu,\mu')&=\frac{(1+\tau)^2}{1+\xi}\sum_{p+q>0}\left(\wxi^p\wxi'\null^q+\wxi^p\wDe_N\wxi'\null^q\wDe_{N^2}\right)+\frac{\xi^2}{1+\xi}\sum_{p+q\geq0}\xi^p\xi'\null^q[\De_NC_{1,(p),N,(q)}+\De_{N^2}C_{1,(p),(q),N^2}]\nonumber\\
&\
  -\De_N\sum_{q\geq1}\wxi'\null^q+\tau'\sum_{q\geq1}\wDe_N\wxi'\null^q\wDe_{N^2}+\text{boundary terms}
\end{align}
whence
\begin{align}
  H(0,\mu')&=\widetilde{H}(0,\mu')/(1+\tau')=\sum_{q>0}\left(\wxi'\null^q-\widetilde{\tau}'\wxi'\null^q\wDe_{N^2}\right)+\text{b.t.}\\
\intertext{and finally}
H(\mu,\mu')&=\frac{1+\tau}{1+\xi}\sum_{p+q>0}\left(\wxi^p\wxi'\null^q+\wxi^p\wDe_N\wxi'\null^q\wDe_{N^2}\right)+\frac{\tau\xi}{1+\xi}\sum_{p+q\geq0}\xi^p\xi'\null^q[\De_NC_{1,(p),N,(q)}+\De_{N^2}C_{1,(p),(q),N^2}]+\text{b.t.}
\end{align}
When $\tau=\tau'$, the formula for $G$ is again recovered. It is interesting
to note that $H$ has here again a Pfaffian structure: if $B$ is the
skew-symmetric $N^2\times N^2$ matrix where all elements but $b_{i,i\+1}=-\tau_i(\tau_i+1)$,
$b_{1,N}=-\tau^2\De_N$, $b_{1,N^2}=-\tau^2\De_{N^2}$, $b_{N,N^2}=-\De_N\De_{N^2}$ are zero, then
\begin{align}
  H(\mu,\mu')&=\frac{1+\tau}{1+\tau+\tau^2}\sqrt{\det(1+BC)}+\text{b.t.}
\end{align}
By induction from the preceding case, we assume that the large deviation function can be computed as if
$\De_N=\De_{N^2}=0$. In that case, we get
\begin{align}
  \log H&\simeq -\demi\sum_{p=1}^\infty\left(-\frac{8}{\pi}\right)^p\frac{1}{p}B_p\\
\end{align}
where $B_p$ is a trace term (we omit the $p$ summation symbols):
\begin{align}
B_1&\simeq\xi_{i_1}\frac{1}{2i_1}\\
  B_p&\simeq\xi_{i_1}\ldots
  \xi_{i_p}\frac{i_1i_2\ldots i_p}{(i_1^2+i_2^2)(i_2^2+i_3^2)\ldots(i_p^2+i_1^2)}
\end{align}

\subsection{General case}

For the general case, where $n_0$ groups of spins are considered
instead of two,
the generalization is quite immediate and similar conclusions hold, that is, $H$
is proportional to a Pfaffian (up to boundary terms) whose leading
term is unchanged  if off-second-diagonal terms are replaced by zero.

As for the computation of $G$, there are two different ways to compute
the leading term of $B_p$. For the inhomogeneous case, it is
convenient to write
\begin{align}
  B_p&\sur{\simeq}{N\rightarrow\infty}p\sum_{i_p=0}^{N^{n_0}-1}\xi_{i_p}\sum_{i_1,i_2,\ldots,i_{p-1}=1}^{i_p}\xi_{i_1}\ldots
  \xi_{i_{p-1}}\frac{(i_1/i_p)(i_2/i_p)\ldots
  (i_{p-1}/i_p)i_p^p}{([i_1/i_p]^2+[i_2/i_p]^2)([i_2/i_p]^2+[i_3/i_p]^2)\ldots(1+[i_1/i_p]^2)i_p^{2p}}\\
&\simeq p\ \sum_{i_p=0}^{N^{n_0}-1}\frac{\xi_{i_p}}{i_p}\int_0^1
  dx_1\int_{0}^1dx_{2}\ldots\int^1_{0}dx_{p-1}\ \ \ \xi_{i_px_1}\ldots
  \xi_{i_px_{p-1}}\frac{x_1x_2\ldots
  x_{p-1}}{(x_1^2+x_2^2)\ldots(x_{p-1}^2+1)(1+x_1^2)}
\end{align}
Let us recall that we consider a particular profile where $\mu_i$ (i.e. $\xi_i$) is constant
in each interval $[0,N-1]$, $[N,N^2-1]$, \ldots, $[N^{n_0-1},N^{n_0}-1]$,
that is $\xi_i=\xi(E[\log(i)/\log(N)])$. Thus,
$\xi_{i_px_1}=\xi(E[\log(i_p)/\log(N)+\log(x_1)/\log(N)])$ and
$\xi_{i_px_1}=\xi_{i_p}$ except for $x_1$ of order $1/N$. As a result,
in this case,
\begin{align}
 B_p &\simeq p\ \sum_{i_p=0}^{N^n-1}\frac{\xi_{i_p}^p}{i_p}\int_0^1
  dx_1\int_{0}^1dx_{2}\ldots\int^1_{0}dx_{p-1}\ \ \ \frac{x_1x_2\ldots
  x_{p-1}}{(x_1^2+x_2^2)\ldots(x_{p-1}^2+1)(1+x_1^2)}\\
&\sim p(\log N)(\xi(0)^p+\xi(1)^p+\ldots+\xi(n-1)^p)\int_0^1
  dx_1\int_{0}^1dx_{2}\ldots\int^1_{0}dx_{p-1}\ \ \ \frac{x_1x_2\ldots
  x_{p-1}}{(x_1^2+x_2^2)\ldots(x_{p-1}^2+1)(1+x_1^2)}
\end{align}
Thus, the coupling vanishes at the level of the large deviation
function, and we have (the vector $\pmb{\mu}=(\mu(1),\ldots,\mu(n_0))$
holds here for the $n_0$ different values of $\mu$ for the $n_0$
groups of spins)
\begin{align}
  g(\pmb{\mu})=\sum_{i=1}^{n_0}\left(\frac{1}{8}-\frac{2}{\pi^2}\left[\text{Arccos}\left(\frac{e^{-2\mu(i)}}{\sqrt{2}}\right)\right]^2\right)
\end{align}
Thus, it appears that the successive groups of spins are
\textit{decorrelated} at the level of the ldf. This decoupling
shows that the situation is here completely different from those
arising in conservative systems, and sheds light on the intimate
relation between the \textit{conservative} character of an observable
and the appearance of severe correlations in a nonequilibrium
stationary state. Here any fluctuation arising somewhere in the system
affects only marginally its vicinity, for this fluctuation is
mainly  locally destroyed by the dissipation.

\section{Conclusion and Perspectives}

We exhibited an example of dissipative system in a stationary state
where a global variable (the cumulative energy $E$) obeys a large
deviation theorem; moreover its large deviation function $f(x)$
appeared to be independent of the injection mechanism. This particular
``boundary layer'' structure made the exact computation of $f$
tractable, and we were also able to compute the large deviation
function associated with the probability of a particular energy
profile. Due to strong inhomogeneities in the stationary state, such a
profile must be defined in a nonstandard way, and there is only one
coherent way to do this. We proved also that the stationary state do
not exhibit correlations as conservative systems do when pulled out of
equilibrium, for the bulk dissipation prevents the fluctuations to be
transported unchanged from one place to another. Incidentally, it
is interesting to mention the reference \cite{basile}, where the
authors come to quite different conclusions : they study the
equilibrium properties of a system driven by Kawasaki+Glauber dynamics
and show that long range correlations do exist, although the system be in equilibrium ; they conjecture thus that
``correlations are more likely to be a generic feature of non
reversible dynamics''. In our case however,  where internal
dynamics are also non reversible, correlations are not induced. Maybe the dynamical details  play an important role, and nonlocal dynamics like Kawasaki's
could impose correlations at the large deviation
level.

 A natural extension of this
work is to add a systematic drift to the system and
test if it would be able to rebuild correlations ; besides, if we
interpret it as a convection phenomenon in a realistic
system, the study of such a competition (and balance) between the dissipation and
the transport in the structuration of the fluctuations could give
interesting insights toward the characterization of dissipative NESS.

But some care must be taken in doing a modification of  internal
dynamics of a dissipative system: actually, it is quite easy to
destroy the large deviation ``structure'' of the energy profile. In
fact, the appearance of strong correlations in a stationary state
seems to be slightly antithetical with a large deviation theorem,
since the latter is mathematically expressed for a sum of $N$
\textit{independent} random variables; if we know that as a rule weak
correlations do not prevent a large deviation theorem to hold, it is
obvious that strongly correlated systems have no reason to verify this
theorem anymore. In this respect, the results of
\cite{seriesofrecentpapers} seem at first sight paradoxical, since
strong correlations are present in their systems; but actually, if their correlations 
are strong in the sense they ``connect'' arbitrary distant places in the
system, they are nevertheless weak as regards their
intensity, since they are inversely proportional to the number of particles.

\section{Appendix} 

The central point of this model is that probabilities $c_{\ldots}$,
whatever the event symbolised by the $\ldots$,  can
be expressed in terms of the $c_{i,j}$ only. 
To see this, let us consider the recursion relation $c_{i,j}$
verifies:
\begin{align}
  c_{i,j}&=\frac{1}{4}\left(\intvidepetit
  c_{i+1,j}+c_{i-1,j}+c_{i,j+1}+c_{i,j-1}\right) \ \ \text{if $i<j$}\label{recrel}\\
c_{0,j}&=1\ \ \forall\ j\geq 0\label{bc1}\\
c_{i,i}&=0\ \ \forall\ i\geq 1\label{bc2}
\end{align}
The equation \myref{recrel} comes from an explicitation of the next
move of the walkers: there are 4 equiprobable moves and for instance
$\text{Prob}[(i,j)|\text{next move is }(i\rightarrow
  i+1)]=c_{i+1,j}$. The boundary conditions \myref{bc2} ensure
that this relation is valid even if $i+1=j$. Similarly the
b.c. \myref{bc1} plays the same role for the event $i\rightarrow i-1$
when $i=1$.

\medskip

Very similar relations can be written for any $c_{i,j,k,\ldots}$ (the
r.h.s of the recursion relation has 6 terms for $c_{i,j,k}$, 8
terms for $c_{i,j,k,\ell}$, etc\ldots). It is then just a matter of
calculation to verify that
the solution of the recursion relation for $c_{i,j,k}$ and
$c_{i,j,k,\ell}$ are
\begin{align}
  c_{i,j,k}&=c_{i,j}+c_{j,k}-c_{i,k}\\
c_{i,j,k,\ell}&=c_{i,j}c_{k,\ell}+c_{i,\ell}c_{j,k}-c_{i,k}c_{j,\ell}
\end{align}
These relations can be generalized to higher orders:
\begin{align}
  c_{i_1,i_2,\ldots,i_{2n+1}}&=c_{i_1,\ldots,
  i_{2n}}-c_{i_1,\ldots,i_{2n-1},i_{2n+1}}+\ldots+c_{i_2,\ldots,i_{2n+1}}\\
c_{i_1,i_2,\ldots,i_{2n}}&=\frac{1}{2^nn!}\sum_{\sig\in
  S_{2n}}\varepsilon(\sig)c_{i_{\sig_1}i_{\sig_2}}c_{i_{sig_3}i_{\sig_4}}\ldots c_{i_{\sig_{2n-1}}i_{\sig_{2n}}}\label{cpairs}
\end{align}
In the last expression, $S_{2n}$ is the set of permutations of
$\{1,2,\ldots,2n\}$ and the $c_{i,j}$ \underline{are antisymmetrised}:
$c_{j,i}\equiv -c_{i,j}$ if $j>i$. The formula \myref{cpairs} show that
the $c_{i_1,\ldots,i_{2n}}$ are Pfaffians associated with the
antisymmetric matrix $C=(c_{i,j})$: as demonstrated in
\cite{derridahakimpasquier}, if $A$ is the $N\times N$ matrix
\begin{align}
  A(\eta,\xi)&=\eta\left(
  \begin{array}{rrrrrr}
    0&1&0& &0&\xi\\
-\!1&0&1&0& &0\\
0&-\!1&0&1&0& \\
&&.&0&.&\\
0&&&.&&1\\
-\xi&0&&&-\!1&0
  \end{array}\right)
\end{align}
We have the relation
\begin{align}\label{det}
  \det\left(\text{Id}+AC\right)=&\left[1-\eta\sum_ic_{i,i+1}+\eta^2\sum_{i<j}c_{i,i+1,j,j+1}-\ldots\right.\nonumber\\
&\ \left.-\xi\left\{\eta
    c_{1,N}-\eta^2\sum_ic_{1,i,i+1,N}+\ldots\right\}\right]^2\\
=&\left[1-\eta C^{(1)}+\eta^2 C^{(2)}-\ldots-\xi\left\{\eta C_{1(0)N}-\eta^2C_{1(1)N}+\ldots\right\}\right]^2
\end{align}

\section{Acknowledgements}

This work greatly benefited from discussions with S. Auma\^\i tre,
T. Bodineau, B. Derrida, S. Fauve,
K. Mallick and F. Werner.

\end{document}

%% file: mesdef.tex
\usepackage{amssymb}
\usepackage{amsmath}

\newcommand{\pa}{\partial}

\newcommand{\myref}[1]{(\ref{#1})}

\newcommand{\de}{\delta}
\newcommand{\De}{\Delta}

\newcommand{\la}{\lambda}

\newcommand{\sig}{\sigma}
\renewcommand{\leq}{\leqslant}
\renewcommand{\geq}{\geqslant}
\newcommand{\lan}{\langle}
\newcommand{\ran}{\rangle}

\newcommand{\demi}{\frac{1}{2}}

\newcommand{\sur}[2]{{\displaystyle\mathop{#1}_{#2}}}

\newcommand{\mcal}[1]{\mathcal{#1}}

\newlength{\somme}
\newlength{\sommep}
\newcommand{\intvide}{\rule[-\sommep]{0cm}{\somme}}

\newlength{\sommebis}
\newlength{\sommepbis}
\newcommand{\intvidepetit}{\rule[-\sommepbis]{0cm}{\sommebis}}

%% file: tracedback.pstex_t
\begin{picture}(0,0)%
\includegraphics{tracedback.pstex}%
\end{picture}%
\setlength{\unitlength}{3315sp}%
\begingroup\makeatletter\ifx\SetFigFont\undefined%
\gdef\SetFigFont#1#2#3#4#5{%
  \reset@font\fontsize{#1}{#2pt}%
  \fontfamily{#3}\fontseries{#4}\fontshape{#5}%
  \selectfont}%
\fi\endgroup%
\begin{picture}(5368,4334)(560,-3648)
\put(5144,-3590){\makebox(0,0)[lb]{\smash{{\SetFigFont{10}{12.0}{\rmdefault}{\mddefault}{\updefault}{\color[rgb]{0,0,0}spin index}%
}}}}
\put(2429,-3590){\makebox(0,0)[b]{\smash{{\SetFigFont{10}{12.0}{\rmdefault}{\mddefault}{\updefault}{\color[rgb]{0,0,0}$j$}%
}}}}
\put(715,-3590){\makebox(0,0)[b]{\smash{{\SetFigFont{10}{12.0}{\rmdefault}{\mddefault}{\updefault}{\color[rgb]{0,0,0}$0$}%
}}}}
\put(715,554){\makebox(0,0)[b]{\smash{{\SetFigFont{10}{12.0}{\rmdefault}{\mddefault}{\updefault}{\color[rgb]{0,0,0}$-t$}%
}}}}
\end{picture}%

%% file: exchange.pstex_t
\begin{picture}(0,0)%
\includegraphics{exchange.pstex}%
\end{picture}%
\setlength{\unitlength}{3315sp}%
\begingroup\makeatletter\ifx\SetFigFont\undefined%
\gdef\SetFigFont#1#2#3#4#5{%
  \reset@font\fontsize{#1}{#2pt}%
  \fontfamily{#3}\fontseries{#4}\fontshape{#5}%
  \selectfont}%
\fi\endgroup%
\begin{picture}(4299,5289)(619,-4888)
\put(2746,-1006){\makebox(0,0)[b]{\smash{{\SetFigFont{11}{13.2}{\rmdefault}{\mddefault}{\updefault}{\color[rgb]{0,0,0}$\Longleftrightarrow$}%
}}}}
\put(2701,-3841){\makebox(0,0)[b]{\smash{{\SetFigFont{11}{13.2}{\rmdefault}{\mddefault}{\updefault}{\color[rgb]{0,0,0}$\Longleftrightarrow$}%
}}}}
\end{picture}%